\documentclass[aip,amsmath,jcp,amsfonts,
preprint,
author-numerical]{revtex4-2}
\usepackage{xcolor}
\usepackage{hyperref}
\usepackage{graphicx}
\usepackage{caption}   
\usepackage{subcaption}  
\usepackage{amssymb}
\usepackage{amsmath}

\graphicspath{{./}}
\begin{document}
	\title{The fluid-lattice gas isomorphism with application to liquid-vapor equilibrium in physisorbed monolayers}
	\author{L.~Shevchenko}
	\email{lshevchenko@mdch1.ukr.education}
	\author{V.~Kulinskii}
	\email{kulinskij@onu.edu.ua}
	\affiliation{Department for Theoretical
		Physics and Astronomy, Odessa National University, Dvoryanskaya 2, 65082 Odessa, Ukraine}
	\begin{abstract}
		Liquid–gas equilibrium is described using a pair of Ising-like thermodynamic variables. The first is a combination of density and energy fields that is even under phase coexistence, derived from the Gibbs–Duhem and Clausius–Clapeyron relations. The second is the normalized density deviation \(M=\rho/\rho_d-1\), which distinguishes the coexisting phases and serves as an odd ordering variable. Combining these variables with the empirical linearities of the binodal diameter and the Zeno line leads to a projective mapping between continuum-fluid and lattice-gas states. The implications of such mapping are tested against coexistence-density and saturation pressure data for the two-dimensional Lennard-Jones fluid, as well as physisorbed molecular monolayers of simple molecular fluids like argon and methane. Finally, a heuristic microscopic interaction-variable formulation that motivates  continuous fluids versus lattice model description is also discussed.
	\end{abstract}
	\maketitle
	\section{Introduction}%
	The Ising model plays an important role in the theory of phase transitions and critical phenomena. Despite its minimalism, it demonstrates many essential aspects of phase coexistence and the critical point, not to mention its various generalizations \cite{book_baxterexact}. Throughout this work, ``Ising model'' is used in this broader symmetry sense and does not necessarily denote the canonical nearest-neighbor Hamiltonian on a particular regular lattice. The lattice geometry may vary, and the spin couplings may extend beyond nearest neighbors or represent effective many-body interactions. The essential requirement is that the zero-ordering field Hamiltonian be invariant under the global transformation $\sigma_i\mapsto-\sigma_i$ of the quasi-spin variable $\sigma_{i} = \pm 1$. It is this $\mathbb Z_2$ symmetry, rather than a specific lattice, coordination number, or interaction range, that supplies the even thermal sector, the odd magnetization-like sector, and the fluid--lattice-gas correspondence developed below. The ``canonical`` textbook ferromagnetic Ising model has the following Hamiltonian:
	\begin{equation}\label{eq:hamis}
		H = -\sum\limits_{\left\langle\, ij \,\right\rangle}\,J_{ij} \, \sigma_{i}\,\sigma_{j} - h \,\sum\limits_{i}\,\sigma_{i}\,,
	\end{equation}
	where $J_{ij}$ is the pair interaction energy, which, in the nearest neighbor approximation, takes an isotropic value $J$,  and $h$ is the external ``magnetic`` field. Using the site occupancy variables $n_{i} = \frac{\sigma_{i}+1}{2}$, one can obtain a lattice gas (LG) representation of the Ising model \cite{book_baxterexact} with the site-site potential $\phi = 4\,J$ and $x = \left\langle n_{i}\right\rangle $ being its density. The Ising model has been remarkably instrumental in determining the singular behavior of thermodynamic quantities in the vicinity of the critical point and in defining universality classes of critical behavior \cite{crit_wilsonkogut_prep1974,crit_fisherrgreview_rmp1998}. The application of this model to a wider range of fluid states is comparatively limited. Obviously, a variety of liquids cannot be described by such a simple model. Besides, the liquid phase has quantitatively different characteristics, both static and dynamic. This originates from the corresponding differences in the density correlation functions of liquid and gas states. The supercritical region has attracted much attention in the search for lines that are considered a natural continuation of the liquid-gas coexistence curve, such as the Widom and Widom-Fisher lines \cite{crit_widom_ptcp1972,fwline_original_jcp1969}. They delineate the boundaries between liquid-like and gas-like behavior in light of density correlation functions \cite{crit_widomline_pnas2005,crit_supercrit_prl2006,crit_widomline_nature2010,liq_brazhkin_pre2012}. However, using density as the most obvious and directly measurable characteristic somewhat obscures the genuine similarity between liquid and gas noted by van der Waals \cite{vdw_nobelecture1910}. The existence of a continuous path that connects any liquid-like and gas-like state, regardless of the specific density, makes Gibbs measures representing such states homotopy equivalent, putting it in proper mathematical terms. This means that from a topological point of view, these states are identical. In contrast, the density-centric description is not invariant and breaks the intrinsic liquid-gas similarity. From this point of view, indeed, many specific liquid-gas division lines can be introduced depending on which specific response function, such as heat capacity, compressibility, etc., is used.
	In the Ising model, or equivalently the lattice gas, the particle-hole symmetry leads to the $\mathbb{Z}_2$ symmetry between states with the same configuration entropy. In view of this fact, to demonstrate the similarity of the liquid and gas states below the critical point, one can use the $\mathbb{Z}_2$-even field, which is a mix of density and entropy, so that the greater density of the liquid state is compensated by its lower entropy, and vise versa for the gas state. Indeed, the existence of such a thermodynamic quantity follows directly from the Clausius-Clapeyron relation 
	\begin{equation}
		\frac{d\mu_{\mathrm{coex}}}{dT} = -\frac{\Delta s}{\Delta\rho}\,.
		\label{eq:CC}
	\end{equation}
	Here $\mu_{coex}$ is the chemical potential on the binodal, $s$ is the entropy density, and $\rho$ is the number density. As a result, one can define a thermodynamic variable:
	\begin{equation}
		\varphi = s + \frac{d\mu_{\mathrm{coex}}}{dT}\,\rho\,.
		\label{eq:symmparam}
	\end{equation}
	Obviously, the entropy contribution, which differs significantly for liquid and gas phases, leads to the asymmetry of the density binodal. The latter can be characterized by its density diameter:
	\begin{equation}\label{eq:diam}
		\rho_d = \frac{\rho_{l}+\rho_{g}}{2}\,,
	\end{equation} 
	where $\rho_{l,g}$ are the densities of the liquid and gas phases. In general, the entropy fluctuations interact with the density field, resulting in the singular temperature behavior of the density diameter \cite{crit_rehrmermin_pra1973,crit_can_diamsing_kulimalo_physa2009,crit_aniswangasym_prl2006}. In the Ising model, the entropy and magnetization (lattice gas density) fields are decoupled due to particle-hole $\mathbb{Z}_2$-symmetry. This leads to a perfectly symmetrical form of the binodal with a straight diameter. It is also observed in many fluid systems in the form of the empirical rectilinear diameter law (RDL)  \cite{crit_diam1}, which provides an effective estimate of the critical parameters \cite{book_frenkelsimul}. If the entropy contribution has little influence on the density field, as in the Sak-Vause lattice model, then the density diameter remains analytic \cite{crit_diamvausesak_jpmathgen1980}. The natural analog of lattice gas density $x$ or Ising-like magnetization $m$ in the context of fluid liquid-gas coexistence is the following variable:
	\begin{equation}\label{eq:xsym}
		x=\rho/(2\rho_d) = \frac{1+m}{2}\,, \quad 0< x < 1.
	\end{equation}
	which can be interpreted as the density of some effective lattice gas model, and the coexisting phases are described via an Ising-like magnetization $\mathbb{Z}_2$-variable $\pm m$. Then, the conjugate field $h_{m}$ variable for such a $\mathbb{Z}_2$-odd order parameter determines the line between liquid and gas states in an ``invariant`` manner as $h_{m} = 0$. A second, closely related empirical regularity is the linearity on the $(T,\rho)$ plane of the Zeno-states that have a unit compressibility factor \cite{eos_zenobenamotz_isrchemphysj1990,eos_zenoapfelbaum1_jpcb2009}:
	\begin{equation}
		Z=P/(\rho\,T)= 1   \,.
		\label{eq:z1}
	\end{equation}
	For a wide class of fluids, this line remains linear far beyond the Boyle temperature and connects smoothly to the low-temperature liquid branch of the binodal. This phenomenological law, augmented with the concept of the liquid-gas state triangle \cite{eos_zenotriapfelbaum_jpchemb2006}, leads to a projective mapping between the asymmetric fluid binodal and the binodal of the lattice gas \cite{eos_zenome0_jphyschemb2010}. It can be considered an extension of the isomorphism between molecular fluid and  the Ising-like model beyond the critical region. We call it fluid-lattice gas global isomorphism, where global refers to the fact that there are correlations among the low-density state (Boyle point), the critical state, the binodal diameter, and the high-density liquid states \cite{eos_zenobenamotz_isrchemphysj1990,eos_zenoapfelbaum1_jpcb2009}. 
	
	The aim of this work is to apply such an approach to 2D systems with the Lennard-Jones interaction potential, in particular to monolayers of simple molecular fluids on a substrate, and to develop an interaction-variable representation that motivates a minimal Ising-like description of their liquid--vapor ordering sector.
	
	The paper is organized as follows. In Section~\ref{sec:gi} we present a simple global isomorphism transformation and its basic corollaries for the liquid-gas equilibrium. Furthermore, in Section~\ref{sec:mono2d} we apply the phenomenological relation for thermodynamic quantities to the analysis of the data on the liquid-gas transition in a 2D Lennard-Jones fluid and monolayers of molecular fluids on a graphite substrate. In Section~\ref{sec:microgi}, we present the microscopic nature of global isomorphism between fluid and lattice Ising-like models. We apply this approach to 2D fluids.  
	\section{Global isomorphism phenomenology}\label{sec:gi}
	The idea of equivalence between gas and liquid equilibrium states is essentially represented by the Ising-like model with particle-hole symmetry. Let $\mu_0(T)$ denote the chemical potential of the global zero-ordering-field manifold. Below the critical point, it is the coexistence chemical potential, $\mu_0(T)=\mu_\mathrm{coex}(T)$ for $T\le T_c$; its canonical supercritical continuation for $T>T_c$ is defined below. Introduce the thermodynamic fields
	\begin{equation}
		h_+ = T,\qquad
		h_- = \mu-\mu_0(T)\,,
		\label{eq:hpm}
	\end{equation}
	then writing
	\begin{equation}
		d\mu = dh_-+\lambda(T)\,dh_+,\qquad
		\lambda(T)=\frac{d\mu_\mathrm{0}}{dT},
		\label{eq:dmuhpm}
	\end{equation}
	and using the Gibbs-Duhem relation, we obtain
	\begin{equation}
		dp=\varphi\,dh_+ + \rho\,dh_-.
		\label{eq:GDPsi}
	\end{equation}
	Using $u=Ts-p+\mu\rho$ and $h_-=\mu-\mu_0(T)$ gives the exact relation
	\begin{equation}
		\Psi_+
		=
		\varphi-\frac{p}{T}+\frac{h_-\rho}{T}.
		\label{eq:PsiPhiRelation}
	\end{equation}
	where $u$ is the energy per unit volume. Therefore, $\Psi_{+}$ is an  energy-like thermodynamic density associated with the thermodynamic variable $\varphi$ and:
	\begin{equation}
		\Psi_{+} = A(T)\,\rho+B(T)\,u\, .
		\label{eq:PsiPlus}
	\end{equation}
	Here:
	\begin{equation}
		A(T) = \lambda(T) - \frac{\mu_{0}}{T}\,,
		\qquad
		B(T) = \frac{1}{T}
		\label{eq:ABcoeff}
	\end{equation}
	In general, $A(T)$ and $B(T)$ depend on $T$ through
	$\mu_\text{0}$ and are continuous functions. If $\mu_\text{0}$ is an analytic function, then so are $A(T), B(T)$. Additionally, it is clear that any analytic function $f(T)$ can be represented as a statistical average of some pure momentum observable:
	$$
	f(T) =  \langle \hat{f}(\mathbf{p})\rangle.
	\label{eq:hmomentum}
	$$
	Therefore, $A(T) = \langle\hat{A}(\mathbf{p})\rangle$ and $B(T) = \langle\hat{B}(\mathbf{p})\rangle$, and we can introduce the corresponding microscopic field:
	\begin{equation}
		\Psi_{+} = \langle\hat\Psi_{+}\rangle,
		\qquad
		\hat\Psi_{+} = \hat{A}(\mathbf{p})\,\hat\rho(\mathbf{r})
		+ \hat{B}(\mathbf{p})\,\hat{u}(\mathbf{r}).
		\label{eq:PsiMicro+}
	\end{equation}
	For a separable classical Hamiltonian in $d$ spatial dimensions, the total energy density is a sum of the ideal and configuration (interaction) terms $u=u_{\rm conf}+ T\rho\,d/2$, where we use energy units for the temperature ($k_{B}=1$). A local microscopic representation of Eq.~\eqref{eq:PsiMicro+} is then
	\begin{equation}
		\hat\Psi_{+}(\mathbf r)
		=\left(\hat A(\mathbf p)+\frac{d}{2}\right)\,\hat\rho(\mathbf r)
		+\hat B(\mathbf p)\,\hat u_{\rm conf}(\mathbf r).
		\label{eq:PsiMicro}
	\end{equation}
	Thus, $\hat \Psi_+$ is a microscopic field that represents a $\mathbb{Z}_{2}$-even property of the coexisting liquid and gas phases. In this respect, its statistical average $\Psi_+$ is analogous to the Ising energy-like thermal operator . The construction of a genuinely odd ordering variable is trivial using the density diameter
	\begin{equation}
		M=\frac{\rho}{\rho_d(T)}-1\,,\qquad -1 < M < 1
		\label{eq:M}
	\end{equation}
	The bound $-1<M<1$ applies to the positive-density coexistence branches. For $T<T_c$, define the spontaneous coexistence magnitude $$M_s(T)=\frac{\rho_l(T)-\rho_g(T)}{\rho_l(T)+\rho_g(T)}>0\,,$$ so that $M_l=M_s$ and $M_g=-M_s$ 
	the subscript $s$ denotes the spontaneous value. 	The field conjugated to $M$ is the correspondingly re-scaled ordering field
	\begin{equation}
		H=\rho_d(T)h_-\,.
		\label{eq:Hnormalized}
	\end{equation}
	Since $\rho=\rho_d(1+M)$, the generating potential for the pair $(H,M)$ is
	\begin{equation}
		\mathcal P(T,H)=p(T,\mu)-H.
		\label{eq:shiftedPressure}
	\end{equation}
	Direct differentiation gives the canonical form
	\begin{equation}
		d\mathcal P=\overline\Psi_+\,dT+M\,dH,
		\qquad
		\overline\Psi_+=\varphi
		-(1+M)\,h_-\frac{d\rho_d}{dT},
		\label{eq:canonicalM}
	\end{equation}
	and therefore
	\begin{equation}
		\overline\Psi_+=\left(\frac{\partial\mathcal P}{\partial T}\right)_H,
		\qquad
		M=\left(\frac{\partial\mathcal P}{\partial H}\right)_T.
		\label{eq:PsiDerivatives}
	\end{equation}
	The canonical pairs are thus $(T,\overline\Psi_+)$ and $(H,M)$. On the coexistence curve $h_-=H=0$, so
	\begin{equation}
		\overline\Psi_+=\varphi=\frac{dp_\sigma}{dT}
		=\Psi_++\frac{p_\sigma}{T}.
		\label{eq:thermalFieldsCoex}
	\end{equation}
	
	The supercritical continuation can now be specified without referring back to $H$ or $M$. For each $T>T_c$, the chemical potential $\mu_0(T)$ is selected directly from the fluid equation of state as the branch issuing continuously from $(T_c,\mu_c)$ on which the isothermal density susceptibility is maximal:
	\begin{equation}
		\left.\frac{\partial^2\rho}{\partial\mu^2}\right|_{T,\mu_0(T)}=0,
		\qquad T>T_c.
		\label{eq:fluidIsingInflection}
	\end{equation}
	because supercritical isotherms clearly have monotonic behavior with the corresponding inflection points. Once $\mu_0(T)$ has been determined independently by Eq.~\eqref{eq:fluidIsingInflection}, the continued diameter is its density–temperature representation,
	\begin{equation}
		\rho_d(T)=\rho\bigl(T,\mu_0(T)\bigr),
		\qquad T>T_c.
		\label{eq:canonicalMuContinuation}
	\end{equation}
	With this continuation, Eqs.~\eqref{eq:hpm}--\eqref{eq:PsiMicro} retain their form above $T_c$ with $\lambda=d\mu_0/dT$.
	The canonical construction, therefore, identifies the zero-ordering-field line:
	\begin{equation}
		H=0:
		\begin{cases}
			T<T_c, & \text{liquid--vapor coexistence with },\, M=\pm M_s(T),\\
			T>T_c, & \text{curve with } M=0.
		\end{cases}
		\label{eq:zeroFieldManifold}
	\end{equation}
	This is the direct fluid analog of the Ising relation $h=0$, $m=0$ above the critical temperature: in the fluid variables, $H=0$ fixes $\mu=\mu_0(T)$, while Eq.~\eqref{eq:canonicalMuContinuation} then gives $M=\rho/\rho_d-1=0$. In the density–temperature plane, the supercritical $M=0$ line is therefore precisely the continuation of the diameter selected by Eq.~\eqref{eq:fluidIsingInflection}.
	At fixed $T$, $\rho_d$ is constant and
	\begin{equation}
		\frac{\partial}{\partial H}
		=\frac{1}{\rho_d}\frac{\partial}{\partial\mu},
		\qquad
		\left(\frac{\partial^2M}{\partial H^2}\right)_T
		=\frac{1}{\rho_d^3}
		\left(\frac{\partial^2\rho}{\partial\mu^2}\right)_T.
		\label{eq:MInflectionFluid}
	\end{equation}
	Consequently, on the selected branch that continued from the critical point,
	\begin{equation}
		H=0
		\quad\Longleftrightarrow\quad
		M=0,
		\qquad
		\left(\frac{\partial^{\,2} M}{\partial H^2}\right)_T=0,
		\qquad T>T_c.
		\label{eq:WidomCanonical}
	\end{equation}
	Here $M_H=(\partial\rho/\partial\mu)_T/\rho_d^2$. Thus, Eq.~\eqref{eq:fluidIsingInflection} is the independent fluid-EoS selection rule, whereas $H=0=M$ and the continued diameter are its representations in the canonical and $(T,\rho)$ variables, respectively. The Ising symmetry applies exactly to the lattice representative and, for a fluid in the Ising universality class, to the leading singular part of the canonical potential in the critical limit:
	\begin{equation}
		\mathcal P_{\rm sing}(T,H)=\mathcal P_{\rm sing}(T,-H),
		\qquad (T,H)\to(T_c,0).
		\label{eq:PevenHypothesis}
	\end{equation}
	If the full canonical potential is approximated by this symmetric representation in the neighborhood of the zero-field line, differentiation gives:
	\begin{equation}
		\overline\Psi_+(T,H)=\overline\Psi_+(T,-H),
		\qquad
		M(T,H)=-M(T,-H),
		\label{eq:Mparity}
	\end{equation}
	which reproduces the Ising parity of entropy and magnetization under magnetic-field reversal. Though Eq.~\eqref{eq:PevenHypothesis} is not imposed globally on the physical fluid potential. Certainly, a real liquid–vapor domain is not invariant under $H\mapsto-H$: the stable liquid branch is bounded at low temperatures by freezing and the triple point, whereas the gaseous branch continues toward the dilute ideal-gas limit. Since our thermodynamic construction uses $\mu_\text{coex}$ and $\rho_d$, the critical region needs separate treatment due to long-range fluctuations leading to nonanalytic behavior of these quantities \cite{crit_rehrmermin_pra1973}. The density field is a composite of basic scaling fields (see, e.g., \cite{book_patpokr}) and the density diameter has the critical $\tau^{1-\alpha}$-singularity due to the linear mixing of even/odd $\mathbb{Z}_2$ fields, where $\tau = 1-T/T_c\,, T<T_c$, $\alpha$ is the specific heat capacity critical exponent. Note that the construction of the thermal field \eqref{eq:PsiMicro+}, which is provided here on a thermodynamic basis, qualitatively corresponds to the result of Bruce\&Wilding \cite{crit_scalfieldsbrucewild_prl1992} in the context of the 2D Lennard-Jones fluid. 
	This construction is the global thermodynamic counterpart of the mixed
	scaling fields introduced by Bruce and Wilding
	\cite{crit_scalfieldsbrucewild_prl1992}.  In their local critical
	description, the energy-like and ordering operators were written as
	\begin{equation}
		\mathcal E_{\rm BW}
		=\frac{u-r_{\rm BW}\rho}{1-s_{\rm BW}r_{\rm BW}},
		\qquad
		\mathcal M_{\rm BW}
		=\frac{\rho-s_{\rm BW}u}{1-s_{\rm BW}r_{\rm BW}}.
		\label{eq:BWoperators}
	\end{equation}
	Our $\mathbb{Z}_{2}$-even density field $\Psi_{+}$ differs from the first of these mixed-operators only by a factor:
	\begin{equation}
		\Psi_+=B(T)\,[u-r_{\rm GI}(T)\rho],
		\qquad
		r_{\rm GI}(T)=-\frac{A(T)}{B(T)}
		=\mu_0(T)-T\frac{d\mu_0}{dT}\,.
		\label{eq:BWmapping}
	\end{equation}
	The relation of the density-based variable $M$ to the mixed ordering operator can be made explicit on coexistence. Let $m_a=\langle\widehat{\mathcal M}_{\rm BW}\rangle_a$ and $m_d=(m_l+m_g)/2$, where $a=l,g$. Here $\widehat{\mathcal M}_{\rm BW}$ is the microscopic operator corresponding to Eq.~\eqref{eq:BWoperators}, $M_a=\rho_a/\rho_d-1$ and $r_{\rm BW}$ and $s_{\rm BW}$ are the Bruce–Wilding mixing coefficients. The Clausius–Clapeyron relation and $u=Ts-p+\mu\rho$ give
	$$
	\Delta u=r_{\rm GI}(T)\Delta\rho,
	\qquad r_{\rm GI}(T)=\mu_0-T\frac{d\mu_0}{dT},
	$$
	and therefore
	$$
	m_a-m_d=\rho_d\,
	\frac{1-s_{\rm BW}r_{\rm GI}(T)}
	{1-s_{\rm BW}r_{\rm BW}}\,M_a.
	$$
	Thus, along coexistence, $M$ is proportional to the centered average of the Bruce–Wilding mixed ordering operator. Hence, $M$ has the same $\mathbb Z_2$-odd parity and a nonzero projection onto the leading magnetization scaling operator, so that the two averages have the same leading behavior $\propto |\tau|^\beta$, where $\beta$ is the order-parameter critical exponent, up to a smooth normalization and subleading corrections. Away from the asymptotic region, the same formal momentum-average construction used in Eq.~\eqref{eq:PsiMicro+} gives
	$$
	M=\langle\widehat M\rangle,\qquad
	\widehat M=\widehat C(\mathbf p)\widehat\rho(\mathbf r)-1,\qquad \langle\widehat C(\mathbf p)\rangle=\rho_d(T)^{-1}.
	$$
	This representation supplies a microscopic observable whose average continues the odd ordering variable through the regular part of the coexistence region. It does not assert an identity between $\widehat M$ and the fluctuating Bruce--Wilding operator, or between their full finite-size distributions; rather, $M$ is a global phenomenological continuation of the same leading odd scaling sector. Since $\alpha = 0$ for the 2D Ising class, the leading correction to the diameter is $\propto \tau \ln \tau$. There can also be a $\tau^{2\beta}$-singular term of the density diameter, known as the Yang-Yang anomaly \cite{crit_yang2_prl1964}, caused by $P-\mu-T$ field symmetry \cite{crit_rehrmermin_pra1973}. It can be described within the Fisher ``complete scaling`` \cite{crit_yydiamfisherorkoulas_prl2000,crit_diamanis_cpl2006}. Such a term should be even more pronounced for $\beta = 1/8$ of the 2D Ising universality class system with $\tau^{1/4}$ diameter singularity if its amplitude is great enough. Direct quantitative tests of singular diameter behavior are rare, particularly for genuinely two-dimensional systems.  To our knowledge, the only such test available is a simulation study of a two-dimensional asymmetric colloid polymer mixture  with a confirmed 2D Ising critical point \cite{crit_diam2dimcolloid_pre2006}. However, no evidence of $t^{1/4}$-contribution in the data has been revealed there. Below, we deal with the data far from the asymptotic region; thus, being incapable of resolving this issue, we leave the question about the diameter singularity beyond the scope of the present paper. But regardless of the specific sub-region of liquid-gas coexistence, whether it is fluctuation or mean-field, we can state that field mixing naturally appears when the $\mathbb{Z}_2$-symmetry invariant description is restored in the entire coexistence region. Thus, we can construct a simplified version of the transformation to the Ising-like variables based on the linear approximation for $\rho_d$, which is a well known empirical fact for simple molecular fluids, and it is widely used for processing liquid-gas coexistence in fluid systems \cite{book_frenkelsimul}. Augmented with another empirical law of the so called Zeno-line linearity \cite{eos_zenobenamotz_isrchemphysj1990}, it leads to a simple mapping between the thermodynamic phase diagram of a simple molecular fluid and that of the lattice gas, proposed in \cite{eos_zenome0_jphyschemb2010}. 
	In view of the definition of $x$ Eq.~\ref{eq:xsym} and the RDL approximation for the density diameter, the corresponding transformation between the fluid density $\rho$ and temperature $T$, and for the LG density $x$ and temperature $t$ takes a projective form:
	\begin{equation}
		\rho = \rho_*\,\frac{x}{1 + z\,\tilde{t}},
		\qquad
		T = T_*\,\frac{z\,\tilde{t}}{1 + z\,\tilde{t}}\,.
		\label{eq:projmap}
	\end{equation}
	Here $\tilde{t} = t/t_c$ and $t_c$ is the critical temperature.

The projective transformation in Eq.~\eqref{eq:projmap} addresses the global asymmetry of liquid--gas coexistence through the rectilinear-diameter approximation and the Zeno-element, or liquid--gas triangle, construction. Its temperature mapping thus has a noncritical geometric origin, whereas mixed-field scaling describes the critical manifestation of the same physical asymmetry. Note that using GI one can relate nonuniversal amplitudes of the Ising model to their fluid counterparts \cite{eos_zenomeglobal_jcp2010}. It is an interesting question for further research to clarify how the Bruce--Wilding mixing coefficients are related to the global isomorphism parameters and how asymptotic mixed-field symmetrization connects to an exact global symmetrization of the liquid--gas coexistence curve.

The parameter $z$ controls the skewness of the density binodal since Eq.~\eqref{eq:projmap} leads to the rectilinear diameter:
	\begin{equation}
		\rho_d^{(\rm lin)}(T)=\frac{\rho_*}{2}
		\left(1-\frac{T}{T_*}\right)\,.
		\label{eq:rdl}    
	\end{equation}
	Note that similar binodal symmetrization procedures were proposed in \cite{eos_zenoapfvorob_lattice2real_jpcb2010,eos_zenovorobsymm_cpl2014}.
	The quantities $(T_*, \rho_*)$ are the parameters of the Zeno-element:
	\begin{equation}
		\frac{T}{T_*} + \frac{\rho}{\rho_*} = 1,
		\label{eq:zenoelement}
	\end{equation}
	which is the image of LG states with $x=1$.  It is a tangent to the extension of the liquid branch of the binodal as $T\to 0$ using the mapping \eqref{eq:projmap}. The scales $T_*$ and $\rho_*$ are estimated from the Boyle/virial construction in the van der Waals (vdW) approximation:
	\begin{equation}\label{eq:tbvdwmy}
		B^{vdW}_2(T_{*}) = 0\,,\quad  T_{*} =  T^{(vdW)}_B  = \frac{a}{b}\,,
	\end{equation}
	and
	\begin{equation}\label{eq:nbvdwmy}
		\rho_*= \frac{ T_* }{B_3\left(\,T_*\,\right)}\,\left. \frac{dB_2}{dT}\right|_{T= T_*}\,.
	\end{equation}
	The parameter $z$ requires a separate input. If $T_c$ is independently known, Eq.~\eqref{eq:critparams} gives $z=T_c/(T_*-T_c)$; alternatively, below we use the interaction-tail scaling estimate $z=d/6$ \cite{crit_globalisome_jcp2010}.
	Here
	\begin{equation}\label{vdw_ab}
		a =\,\, -\frac{S_{d}}{2}\,\int\limits_{\sigma}^{+\infty}\Phi_{attr}(r)\,r^{d-1}\,dr
	\end{equation}
	and $\Phi_{attr}(r)$ is the attractive part of the interaction potential $\Phi(r)$ that determines the corresponding virial coefficients $B_2,\,B_3$; $\sigma$ is the effective diameter of a particle, such that $b = \frac{V_{d}}{2}\,\sigma^{d}$ and $S_d, V_{d}$ are the area and volume of a unit $d$-sphere, respectively. In physical terms, $\rho_{*}$ corresponds to a fully occupied state $x=1$ of the lattice gas (see also \cite{eos_zenosanchez_jpcb2016} where $\rho_{*}$ was related to the hypothetical disordered liquid (glass) state at zero temperature). This defines the number of cells  $N_{\text{cells}} = \rho_{*}\,V$ for the isomorphic  lattice gas model in a volume $V$ of a fluid. In terms of global isomorphism with the corresponding LG (Ising-like) model, the low-temperature states $\rho \to 0$ and $\rho \to \rho_*$ inherit the duality of the LG states $x \to 0$ and $x \to 1$. From the physical point of view, the state $x\to 1$  can be treated as an almost ideal gas of unoccupied sites. Thus, the high-density liquid state can be considered a gas of voids. They interact through the same site-site attractive potential, which sets the value of $T_*$. 
	The critical temperature and density of a fluid, according to \eqref{eq:projmap}, are:
	\begin{equation}
		\rho_c = \frac{\rho_*/2}{1+z},
		\qquad
		T_c = T_*\frac{z}{1+z},
		\label{eq:critparams}
	\end{equation}
	which implies the testable relation on the critical parameters:
	\begin{equation}
		2\,\frac{\rho_c}{\rho_*} + \frac{T_c}{T_*} = 1.
		\label{eq:testrel}
	\end{equation}
	For those fluids where the LRD is observed in a wide region of liquid-gas coexistence, Eq.\eqref{eq:testrel} is supported by the experimental data \cite{eos_globiso_cmp2025}. From \eqref{eq:projmap} and standard thermodynamic relations
	\begin{equation}
		\rho =
		-\frac{1}{V}
		\left(\frac{\partial\, \Omega_{F}}{\partial \mu}\right)_T,
		\qquad
		x
		=
		-\frac{1}{N_{\text{cells}}}
		\left(\frac{\partial\, \Omega_{\text{LG}}}{\partial \nu }\right)_t,
	\end{equation}
	with $\Omega = -p \,V$ and $\nu$ as the LG chemical potential, the following relation between the pressures of these systems can be obtained:
	\begin{equation}
		p_{F}(\mu,T)= \rho_* p_{\text{LG}}(\nu(T,\mu),t(T))+C(T,\mu)\,.
		\label{eq:pfplg}
	\end{equation}
	Here $C(T,\mu)$ is an analytic function across coexistence.  The projective mapping \eqref{eq:projmap} fixes the remaining linear normalization transverse to the coexistence curve as
	\begin{equation}
		2h\equiv\nu(T,\mu)-\nu_0(t(T))
		=\frac{\mu-\mu_{\rm coex}(T)}{1+z\tilde{t}(T)}
		+o\!\left([\mu-\mu_{\rm coex}(T)]\right)\,.
		\label{eq:chemicalPotentialMap}
	\end{equation}
	To preserve the absolute saturation-pressure correspondence and to ensure that the analytic background does not alter coexistence densities, we impose the following conditions:
	\begin{equation}
		C\bigl(T,\mu_{\rm coex}(T)\bigr)=0,
		\qquad
		\left.\left(\frac{\partial C}{\partial\mu}\right)_T
		\right|_{\mu=\mu_{\rm coex}(T)}=0,
		\qquad T\leq T_c.
		\label{eq:CcoexZero}
	\end{equation}
	Hence, the coexistence result is fixed by the singular pressure structure, while any continuation away from coexistence is a specific LG model of choice encoded in $\nu(T,\mu)$ and $C(T,\mu)$. Naturally, the pressure of the LG goes to infinity $p_{LG}\to \infty$ as $x\to 1$. But this divergence is just the mirror of the divergence of the Ising model free energy for the dual state ($x\to0$) with all spins equal $-1$ ($h\to -\infty$). Clearly, the function $C(T,\mu)$ must cancel such divergence in order to restore the corresponding ideal gas asymptotic behavior for $p_{F}$:
	$$p_{F} \to \rho \,T\,,\quad \rho \to 0 $$ 
	Further, we restrict our consideration of the experimental data in the coexisting region where $C=0$. 	In particular, the critical compressibility ratio $Z$ plays an important role in thermodynamics. It is a dimensionless marker for classes of similar thermodynamic behavior, and therefore, it can be used to establish the relation between real and model fluid systems.  Eq.~\eqref{eq:pfplg} and \eqref{eq:CcoexZero} lead to the following relation between the critical compressibility factors of these systems. Indeed, the compressibility factors are:
	\begin{equation}
		Z_{F} = \frac{p_{F}}{\rho\, T}\,, \quad Z_{LG} = \frac{p_{LG}}{x\, t}
		\label{eq:z}
	\end{equation}
	so from Eq.~\eqref{eq:pfplg} with account of Eq.~\eqref{eq:CcoexZero}, for the compressibility factor value at the critical point, we get:
	\begin{equation}
		Z^{(c)}_{F} = \frac{(1+z)^2}{z}\,\frac{t_c}{\,T_* }\,
		Z^{(c)}_{LG}
		\label{eq:zcrelation}
	\end{equation}
	Note that due to Eqs.~\eqref{eq:critparams},\eqref{eq:pfplg} $\rho_*$ cancels out. 
	This relation has been shown to be consistent with the available data for 3D fluid systems \cite{eos_zenozcassocme_jcp2014}. For the 2D systems, the relation \eqref{eq:zcrelation} can be tested using experimental data and information on the exact solutions of relevant Ising-like models.
	\section{Application to 2D monolayers of model and real fluids}\label{sec:mono2d}
\begin{figure}[h!]
		\begin{subfigure}{0.48\textwidth}
			\centering
			\includegraphics[width=\linewidth]{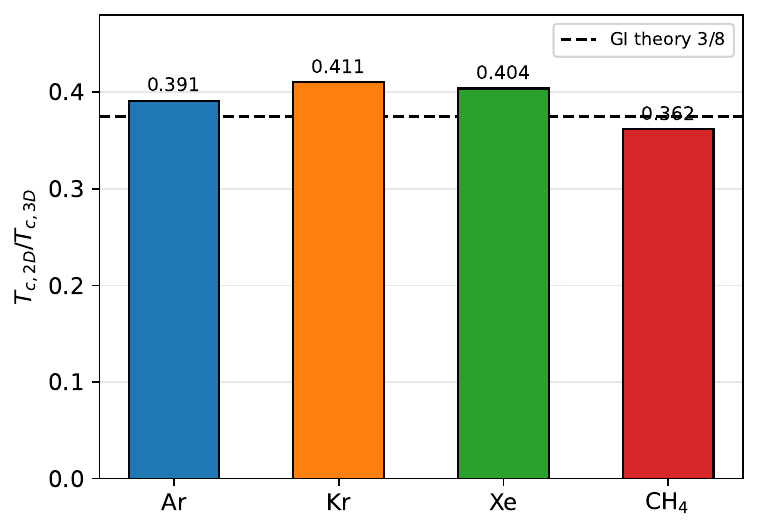}
		\end{subfigure}
		\hfill
		\begin{subfigure}{0.48\textwidth}
			\centering
			\includegraphics[width=\linewidth]{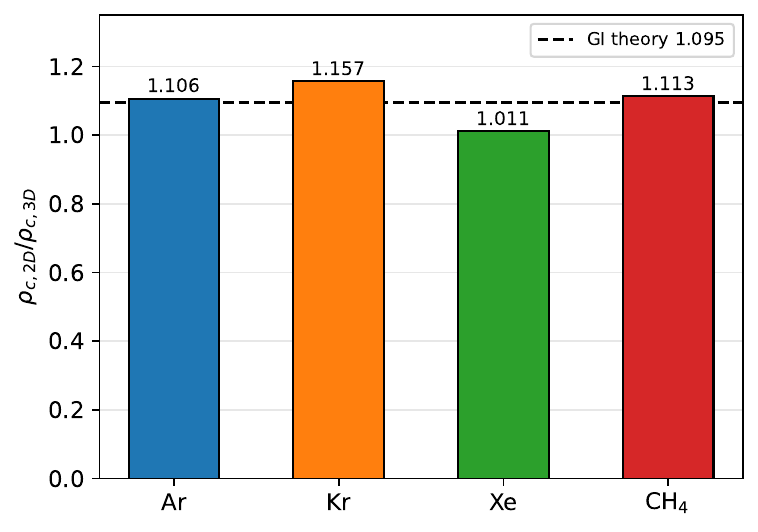}
		\end{subfigure}
		\caption{Comparison between the experimental data from Table~\ref{tab:z2dmono} and the theoretical prediction \eqref{eq:ratio2d3d} (dashed lines).}
		\label{fig:ratios}.
	\end{figure}

\begin{figure}[h!]
		\centering
		\includegraphics[width=0.8\textwidth]{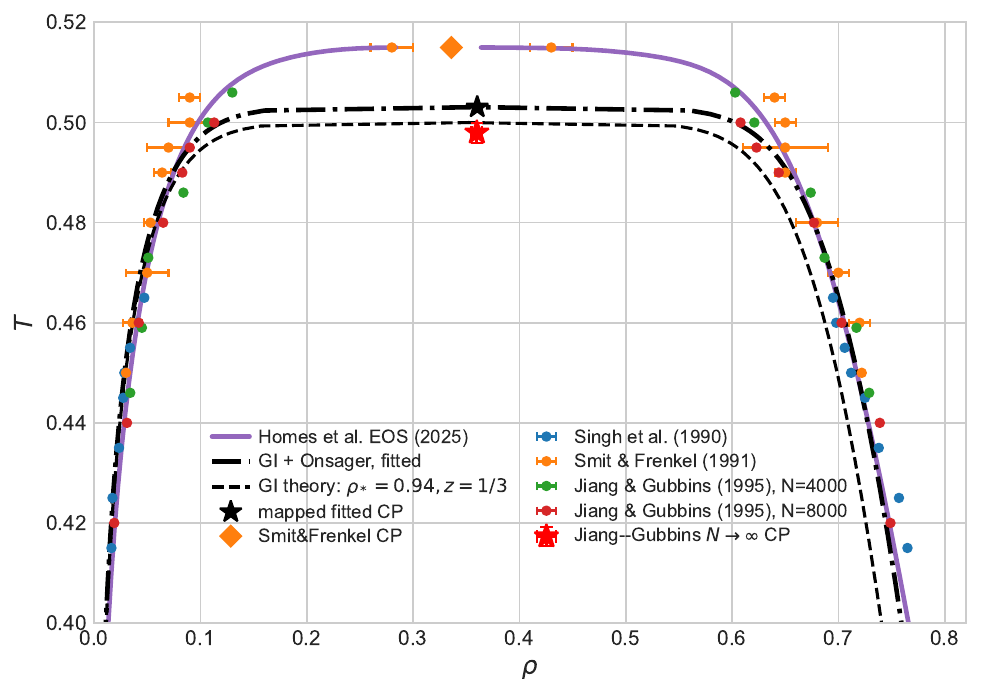}
		\caption{Binodal data simulations for 2D LJ fluid. Solid curve is the binodal based on the 18-parametric analytic EoS of \cite{liq_2dljfluid.jml2025} tuned to the Smit\&Frenkel critical temperature $T_c = 0.515$ \cite{crit_lj2dimsmitfrenkel_jcp1991} and fitted $\rho_{c} = 0.355$. Dashed line is the Onsager binodal mapped with theoretical values \eqref{eq:ztrho2d}, dashed-dotted curve uses fitting parameters $\rho_{*} =0.963, z=0.335$, corresponding to the critical parameters $\rho_c = 0.36\,, T_c = 0.503$}\label{fig:2dljbin}
	\end{figure}

\begin{figure}[h!]
		\centering		\includegraphics[width=0.8\textwidth]{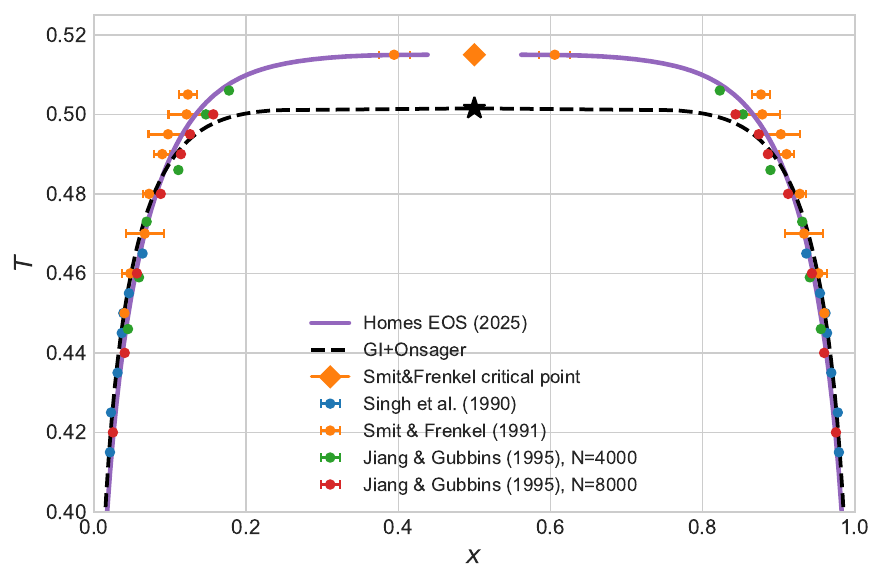}
		\caption{Symmetrized binodal data with Onsager's binodal mapped with $z = 0.335$ and corresponding $T_c = 0.502$}\label{fig:2dljbinsymm}
	\end{figure}

\begin{figure}[h!]
		\centering
		\includegraphics[width=0.8\textwidth]{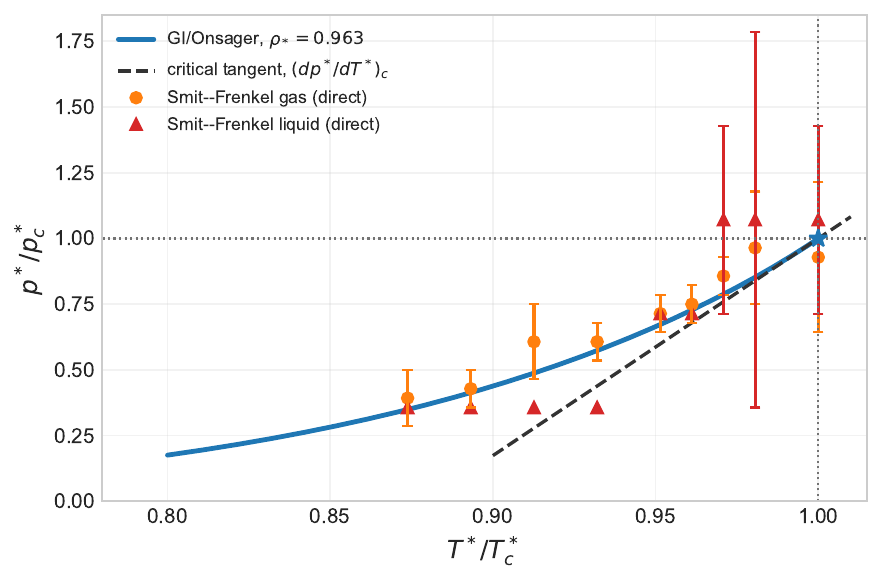}
		\caption{GI/Onsager saturation pressure \eqref{eq:pfreeis} compared with
			the Smit\&Frenkel~\cite{crit_lj2dimsmitfrenkel_jcp1991} simulation data for gas and liquid pressures.}	\label{fig:lj2dPressureTemperatureGI}
	\end{figure}

\begin{figure}[h!]
		\centering	\includegraphics[width=0.8\linewidth]{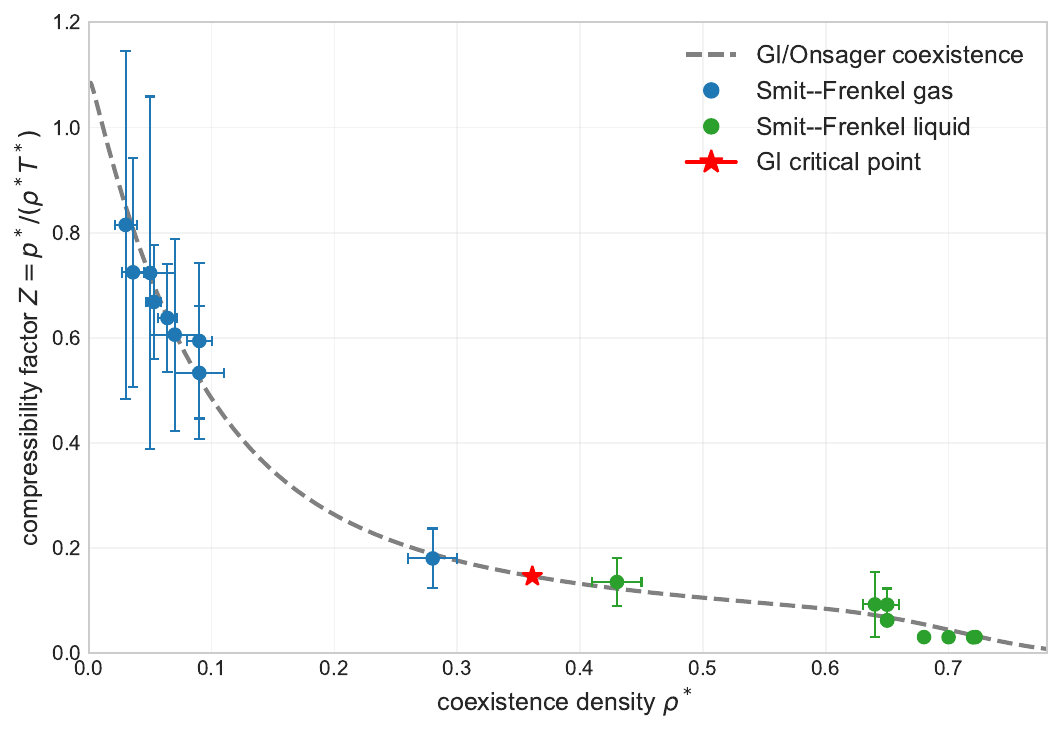}
		\caption{The compressibility factor $Z$ along the liquid-gas coexisting curve of 2D LJ fluid. The points are the simulation data \cite{crit_lj2dimsmitfrenkel_jcp1991}.}
		\label{fig:z2dljdata}
	\end{figure}

\begin{figure*}[h!]
		\begin{subfigure}{0.48\textwidth}
			\centering
			\includegraphics[width=\linewidth]{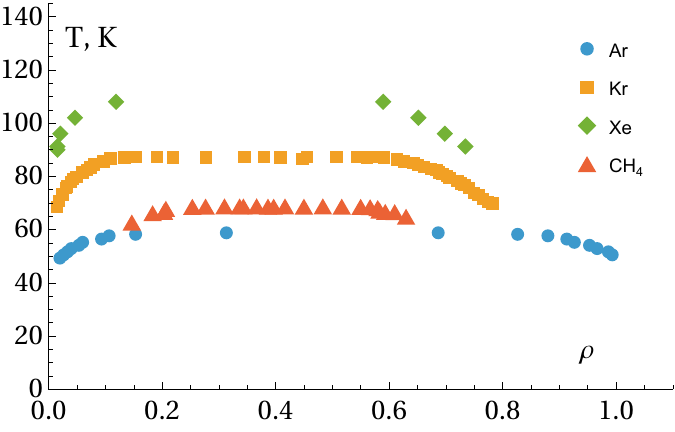}
			\caption{}
			\label{fig:fig_1fluidata2d}
		\end{subfigure}
		\hfill
		\begin{subfigure}{0.48\textwidth}
			\centering			\includegraphics[width=\linewidth]{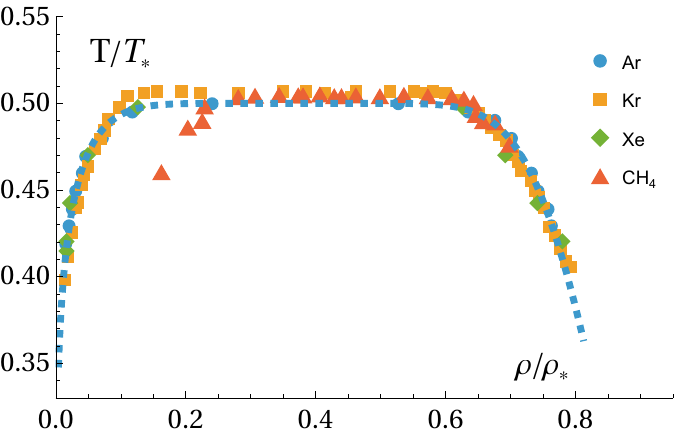}
			\caption{}
			\label{fig:fig_2fluidata2d}
		\end{subfigure}
		\caption{Binodal data of MC simulations for $Ar,Xe$ \cite{eos_2darxe_condmat2012} and experimental data for $CH_4$ \cite{eos_2dabsrbch4_prb1986} in original physical units (left);  the same data scaled to $T_*\,,\rho_*$ (right). Dashed curve is the Onsager's solution mapping according to \eqref{eq:projmap}.}
		\label{fig:binodaldata}.
	\end{figure*}

	There are a number of numerical simulations of 2D fluid  \cite{crit_lj2dim_jcp1990,crit_lj2dimsmitfrenkel_jcp1991,crit_lj2d_molphys1995} with the Lennard-Jones (LJ) interaction potential
	\begin{equation}
		\label{eq:lj}
		\Phi_{LJ}(r)=4\,\varepsilon\, \Big(\Big(\frac{\sigma}{r} \Big)^{12} -\Big(\frac{\sigma}{r} \Big)^6 \Big)\,.
	\end{equation}
	The liquid-gas transition is clearly observed here. These data are presented in Fig.~\ref{fig:2dljbin}. The planar character of such systems allows for the use of Onsager's exact solution of the Ising model to obtain relevant thermodynamic information for these systems based on the global isomorphism relations obtained in the previous section. We represent Onsager's exact solution for the binodal in LG terms:
	\begin{equation}\label{eq:isingbinodal2}
		x = 1/2\left(1\pm f(t)^{1/8}\right)\,,\quad f(t) =
		1-\frac{1}{\sinh^4\left(\,2J/t \,\right)}\,,
	\end{equation}
	and apply Eq.~\eqref{eq:projmap} to map it to the data. The reduced Lennard–Jones variables are $T^*=T/\varepsilon$, $\rho^*=\rho\sigma^d$, and $p^*=p\sigma^d/\varepsilon$, with $k_B=1$ and $d=2$ for a monolayer. Superscript stars indicate reduced units, whereas the subscripted $T_*$ and $\rho_*$ denote the mapping scales. Below, we omit superscript stars when quoting reduced numerical values. Also, we assume that the LJ parameter $\varepsilon$ can be identified with the interaction constant of the lattice gas $\varepsilon \equiv \phi$ based on the fact that $\varepsilon$ is the interaction energy at an equilibrium distance. Equations~\eqref{eq:tbvdwmy} and \eqref{eq:nbvdwmy} supply $T_*$ and $\rho_*$. Separately, the scaling estimate $z=d/6$ for an $r^{-6}$ attractive tail gives $z=1/3$ in two dimensions \cite{crit_globalisome_jcp2010}. Together, these choices yield:
	\begin{equation}
		z=1/3\,, \quad T_* = 2\,, \,\quad \rho_{*} \approx 0.941
		\label{eq:ztrho2d}
	\end{equation}
	which yields magnitudes of the critical parameters for the 2D LJ fluid presented in Table~\ref{tab:2dljcrit}. There is an obvious trend in the finite-size estimate of the critical temperature $T^{(N)}_c \to T^{(\infty)}_{c} \approx 0.50$. For a detailed discussion of finite-size effects, we refer to \cite{crit_lj23dfinitesize_jthrmphys1994,crit_lj2d_molphys1995}. As long as the truncated potential is used in MC simulations, affecting the critical point locus, we can use $\rho_{*}$ and $z$ as fitting parameters. Naturally, $\rho_{*}$-fitting improves the correspondence with the liquid branch, while $z$ does not  differ considerably from the theoretical value. For comparison, we also provide the exact density symmetrization, with no need for fitting $\rho_*$ (see Fig.~\ref{fig:2dljbinsymm}). The only fitted parameter is $z$, which determines $T_c$ in accordance with \eqref{eq:critparams}. Overall, we observe quite a reasonable agreement with the data. Also, using Eq.~\eqref{eq:critparams}, it is easy to relate the monolayer critical temperature and density to the values for the bulk phase of the LJ fluid. For 3D LJ, we have \cite{crit_globalisome_jcp2010}:
	\begin{equation}
		z=1/2\,,\quad T_* = 4\,\,,\quad \rho_{*} \approx 0.967.
		\label{eq:ztparam2d}
	\end{equation}
	As a result, we get the following simple relations:
	\begin{equation}
		T^{(2D)}_{c} \approx 3/8\,T^{(3D)}_{c}\,, \quad \rho^{(2D)}_{c} \approx 1.095\,\rho^{(3D)}_{c} 
		\label{eq:ratio2d3d}
	\end{equation}
	which can be tested using the available data. We summarize the critical parameters in Table~\ref{tab:z2dmono} and in Fig.~\ref{fig:ratios}. As we can see, there is a reasonable agreement with experimental values. 
	\begin{table}[htb]
		\caption{Critical parameters for monolayers modeled as 2D LJ fluid.}
		\label{tab:z2dmono}
		\begin{center}
			\begin{tabular}{|c|c|c|c|c|}
				\hline
				&$Ar$& $Kr$ & $Xe$ & $CH_4$\\
				\hline
				$ 3D: T_c^{(\text{exp})}, K\,$ \cite{book_nist69} & $150.8$& $209.48$ & $289.7$ & $190.6$\\
				\hline
				$ 3D: \rho_c^{(\text{exp})}\,$\cite{book_nist69} & $ 0.317$ & $ 0.304$& $ 0.348$&$ 0.317$\\
				\hline
				\hline
				$ 2D: T_c^{(\text{exp})}, K$ & $59$\cite{eos_2dmononoble_physrev1979}& $86$\cite{eos_2dmononoble_physrev1979}& $117\cite{eos_2dmononoble_physrev1979}$ &$69$ \cite{eos_2dabsrbch4_prb1986}\\
				\hline
				$ 2D: \rho_c^{(\text{exp})}$ & $ 0.357$\cite{eos_2dmono_jcp2018} & $ 0.358$\cite{eos_2dmono_jcp2018}& $ 0.344$\cite{eos_2dmono_jcp2018}&$  0.368$\cite{crit_lj2d_molphys1995}\\
				\hline
				\hline
				$ 2D: T_c^{(GI)}, K$ & $57$& $79$& $109$ & $71$\\
				\hline
				$ 2D: p_c^{(GI)}, 10^{-3} N/m$ & $0.37 $& $ 0.44$& $0.47$ & $ 0.38$\\
				\hline
				$2D: \rho_c^{(\text{GI})}$ 
				& \multicolumn{4}{c|}{$0.353$} \\
				\hline
				$3D: \rho_c^{(\text{GI})}$ 
				& \multicolumn{4}{c|}{$0.322$} \\
				\hline
			\end{tabular}
		\end{center}
	\end{table}
	\begin{table}
		\centering
		\begin{tabular}{|c|c|c|c|c|c|c|c|}
			\hline
			\hspace{0.2cm} LJ ``6-12`` fluid \hspace{0.2cm}&
			\hspace{0.2cm} $T_c$ \hspace{0.2cm}& \hspace{0.2cm}$\rho_c $\hspace{0.2cm}& \hspace{0.2cm}$Z_c$\hspace{0.2cm}\\
			\hline
			GI &0.5&\hspace{0.2cm} 0.354 \hspace{0.2cm}&\hspace{0.2cm} 0.146 \hspace{0.2cm}\\
			\hline
			\hspace{0.2cm}		Jiang\&Gubbins, N=8000  \cite{crit_lj2d_molphys1995} \hspace{0.2cm}&\hspace{0.2cm} $0.498 \pm 0.002$ \hspace{0.2cm}&\hspace{0.2cm}$ 0.360 \pm 0.005$ \hspace{0.2cm}& - \hspace{0.2cm}\\
			\hline
			\hspace{0.2cm}		Smit\&Frenkel, N=512 \cite{crit_lj2dimsmitfrenkel_jcp1991} \hspace{0.2cm}&\hspace{0.2cm} $0.515 \pm 0.002$ \hspace{0.2cm}&\hspace{0.2cm}$ 0.355 \pm 0.003$ \hspace{0.2cm}& - \hspace{0.2cm}\\
			\hline
		\end{tabular}
		\caption{The critical point parameters of 2D $LJ$ fluid according to \eqref{eq:critparams} and \eqref{eq:ztrho2d} in comparison with the results \cite{crit_lj2dimsmitfrenkel_jcp1991,crit_lj2d_molphys1995}.}\label{tab:2dljcrit}
	\end{table}
	Using the relationship between the grand potential and pressure:
	\begin{equation}
		p_F = - \Omega_F/V\,,
	\end{equation}
	we can get the fluid pressure  along the coexistence line:
	\begin{equation}
		p_F (T)
		=
		-\rho_*\, F_{\text{Is}}(t(T)) - \rho_* \frac{q\,J}{2}\, ,
		\label{eq:pfreeis}
	\end{equation}
	where $q$ is the coordination number of the lattice and $J$ (e.g., for the square lattice $q=4$). Besides, due to the continuity of the $dp_{F}/dT$ along the coexistence line with account of Eq.~\eqref{eq:CcoexZero} we can derive the critical slope:
	\begin{equation}
		\left. \frac{dp_{F}}{d\,T}\right|_{T_c} =    
		\rho_* \frac{t_c(1+z)^2 }{z\, T_*}\, s_{\text{Is}}(t_c)\,, \qquad s_{\text{Is}}  = - \frac{d\,F_{\text{Is}}(t(T)) }{d\,t}
	\end{equation}	
	For the square-lattice model in the reduced units used here, $\varepsilon=\phi=4J=1$, and therefore
	$t_c=[2\ln(1+\sqrt{2})]^{-1}=0.567\ldots$. The exact critical entropy per site is \cite{book_baxterexact}
	\begin{equation*}
		s_{\rm Is}(t_c)=\frac{1}{2}\ln 2+\frac{2G}{\pi}
		-\frac{\ln(1+\sqrt{2})}{\sqrt{2}}=0.306\ldots,
	\end{equation*}
	where $G$ is Catalan's constant. Using the two-dimensional parameters in Eq.~\eqref{eq:ztrho2d} we get
	\begin{equation*}
		\left.\frac{dp_F}{dT}\right|_{T_c^-}
		\approx 0.436\,.
	\end{equation*} 
	The direct comparison between \eqref{eq:pfreeis} and the saturation pressure data of Smit\&Frenkel is shown in Fig.~\ref{fig:lj2dPressureTemperatureGI} (big uncertainties for gas low pressure data are not shown).
	As the relation \eqref{eq:zcrelation} directly follows from \eqref{eq:projmap} and \eqref{eq:pfplg}, we also represent the data on 2D LJ fluid for the compressibility factor $Z$ depending on the density $\rho$ along the coexistence curve. Obviously, it is a monotonous function along the binodal. The result is presented in Fig.~\ref{fig:z2dljdata}.
	Now we process the binodal data for the fluid monolayers on a substrate and compare it with that of the 2D LJ fluid. We collected the data from both numerical simulations \cite{crit_lj2dim_jcp1990,crit_lj2dimsmitfrenkel_jcp1991,crit_lj2d_molphys1995} and 2D monolayers of various molecular fluids  \cite{eos_2dabsrbch4_prb1986,eos_2darxe_condmat2012}. The results are shown in Fig.~\ref{fig:binodaldata}. Clearly, these fluids belong to the same thermodynamic similarity class, which is supported by Fig~\ref{fig:fig_2fluidata2d}. Only the gas branch data of $CH_4$ from \cite{eos_2dabsrbch4_prb1986} appear to exhibit irregular behavior. This raises some reservations, as noted in \cite{crit_lj2d_molphys1995}, caused by additional absorption due to surface inhomogeneities. For the symmetrization procedure with methane, the liquid branch data are sufficient. 
	For comparison, the same data were symmetrized, and the effective temperature variable $\tilde{t}$ in Eq.~\eqref{eq:projmap} is used at $T_* = 2$. This allowed for a direct comparison with the theoretical curve of the Ising model \eqref{eq:isingbinodal2}. As one can see, the relative error values do not exceed $5\,\%$. Note that for a $2D$ system with a short-ranged potential, the extreme flatness of the binodal dome is a major problem for an analytic EoS, where the resolution comes at the expense of using many fitting parameters \cite{eos_2dlj_canjphys1986,liq_2dljfluid.jml2025}. The recent Helmholtz EOS of Homes et al.~\cite{liq_2dljfluid.jml2025} provides a useful independent thermodynamic representation, but its critical locus must be interpreted carefully.  In constructing that EOS, the authors used the Smit–Frenkel value $T_c=0.515$. Consequently, its $T_c$ is rather an estimate of the finite-size trend (see \cite{crit_lj2d_molphys1995}).  Moreover, since the residual Helmholtz energy is analytic at the imposed critical point, its asymptotic coexistence width has the  mean-field form, rather than the exact 2D-Ising one with $\beta = 1/8$. The EOS can consequently be accurate over a broad thermodynamic domain, while the direct Onsager-based symmetrization gives the correct critical-region shape.
	We did not add 2D pressure values for real fluids here, as the majority of experimental data are represented in terms of graphical adsorption isotherms for 3D vapor pressure above the surface. They need to be deciphered and recalculated to obtain $2D$ spreading pressure of a layer using the Gibbs adsorption equation \cite{eos_2dmono_rpm2007}, and we plan to do it as separate work.
	\begin{figure}[h!]
		\begin{subfigure}{0.46\textwidth}
			\centering
			\includegraphics[width=\linewidth]{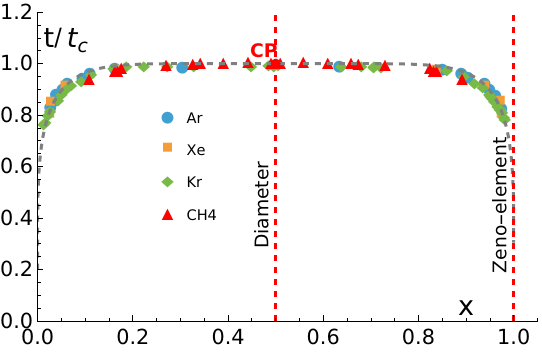}
		\end{subfigure}
		\hfill
		\begin{subfigure}{0.46\textwidth}
			\centering			\includegraphics[width=\linewidth]{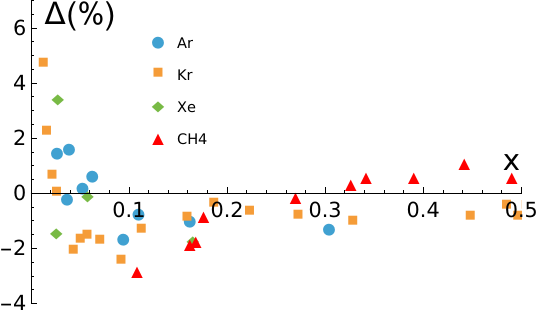}
		\end{subfigure}
		\caption{Symmetrized binodal data and Onsager's binodal.}
		\label{fig:binodalsumdata}.
	\end{figure}
	\section{Heuristic microscopic construction of fluid-lattice gas isomorphism}\label{sec:microgi}%
	The construction of the global isomorphism transformation \eqref{eq:projmap} mirrors the empirical facts of the RDL, the Zeno-line linearity \cite{eos_zenobatschinski_annphys1906,eos_zenobenamotz_isrchemphysj1990,eos_zenoapfelbaum1_jpcb2009}, and the liquid-gas triangle construction \cite{eos_zenotriapfelbaum_jpchemb2006}. As noted above, it also reflects the ''particle-hole'' Ising-model symmetry, which is distorted by the use of the density variable as the main configuration field on which the Hamiltonian depends.  As has been shown above, such a symmetry can be restored in terms of the composite field. Therefore, the derivation of \eqref{eq:projmap} from the microscopic principles becomes important. In this section, we present the microscopic approach to the phenomenology of global isomorphism. 
	
	Let us consider the starting point of statistical theory — the  partition function:
	\begin{equation}
		\Xi_{fluid} =\Xi^{(id)}\,\Xi^{(int)}_{N}
	\end{equation}
	with the corresponding interaction contribution:
	\begin{equation}
		\Xi^{(int)}_{N} =\int\limits_\Lambda e^{-\beta \sum\limits_{i<j} \Phi(r_{ij})}\,d\Gamma_N \,,\quad  d\Gamma_N = \frac{1}{V^N} \prod_i d\mathbf{r}_i .
	\end{equation}
	All the information about particle configurations is encoded in the interaction values of $\Phi(r_{ij})$, and the particle positions $\mathbf{r}_i$ in such a case become redundant. Clearly, only $\sim d\,N$ of $N(N-1)/2$ functions $\Phi(r_{ij})$ are functionally independent ($d$ is the number of dimensions) due to obvious constraints $r_{ij} = |\mathbf{r}_{ik} - \mathbf{r}_{jk}|$. Therefore, we can introduce the corresponding distribution function:
	\begin{equation}
		p_N ([\mathcal{J}];\rho) = \int \prod_{i<j} \delta\left(\Phi(r_{ij}) - \mathcal{J}_{ij}\right) \, d\Gamma_N \,,\label{eq:pn}
	\end{equation}
	and 
	\begin{equation}
		\Xi^{(int)}_{N}=\int e^{-\beta\sum\limits_{i<j}\mathcal{J}_{ij}}\,p_N([\mathcal{J}];\rho)\prod\limits_{i<j}d\mathcal{J}_{ij}.
		\label{XiJform}
	\end{equation}
	Note that $p_N$ does not depend on temperature but only on $V$ and $N$. The $\mathcal{J}$-distribution density $p_N $ is a general form of the cumulant expansion \cite{book_yukhgol_en}:
	$$
	p_N([\mathcal{J}]; \rho)=\int d\omega\;
	\exp\left(2\pi i\,\sum_{a} \omega_a \mathcal{J}_a + \sum_{n=1}^{\infty}
	\frac{(2\pi i)^n}{n!}
	\sum_{a_1,...,a_n} \mathfrak{M}_n^{a_1\dots a_n}
	\omega_{a_1}\cdots\omega_{a_n}
	\right).
	$$
	This expression encodes all statistical information about the random variables $\mathcal{J}_a$ in terms of cumulants $\mathfrak{M}_n$:
	$$
	\begin{aligned}
		\mathfrak{M}_1 &= \langle \Phi(r_{ij}) \rangle, \\
		\mathfrak{M}_2 &= \langle \Phi(r_{ij})\Phi(r_{kl}) \rangle - \langle \Phi(r_{ij}) \rangle \langle \Phi(r_{kl}) \rangle, \\
		\ldots
	\end{aligned}
	$$
	Generally, we can express $p_{N}$ in the form:
	\begin{equation}
		p_N([\mathcal{J}];\rho) = \exp\left( 
		\sum_{i<j}\Gamma_{1}\mathcal{J}_{ij} +  \frac{1}{2!}\sum_{i<j}\sum_{k<l} \Gamma_{2} \mathcal{J}_{ij} \mathcal{J}_{kl} + \dots
		\right).
	\end{equation}
	The linear term appears due to the averaging of the $\mathcal{J}$-distribution. As a result, we obtain the effective $\mathcal{J}$-Hamiltonian: 
	\begin{equation}
		\beta\,\mathcal{H}_{eff}([\tau]) = \beta \,\mathcal{H}_{id} + \left(\beta - \Gamma_{1} \right)\,\sum_{i<j}  \mathcal{J}_{ij}.
		\label{eq:Heff}
	\end{equation}
	Here $\Gamma_1$ is some function of the density $\rho=N/V$. Such a representation demonstrates  that one can consider the ``Zeno-line`` states:
	\begin{equation}
		\beta = \Gamma_1(\rho)
		\label{eq:zeno0}
	\end{equation}
	where the interaction is ``switched off`` as all the other nonlinear terms are due to a pure coordinate transformation of configuration space. Taking into account that $\Gamma_{1} \to \beta_* \ne 0$ as $\rho \to 0$, we can relate $\beta_*$ to $T_* = 1/\beta_*$ - the Boyle temperature in the vdW approximation  when all other density-dependent terms in \eqref{eq:pn} are negligible. Therefore, we may redefine the temperature parameter
	\begin{equation}
		\tilde{\beta} = \beta - \beta_* >0
		\label{eq:betabeta}
	\end{equation}
	which, up to the proper scale, is similar to the temperature transformation which was introduced in \eqref{eq:projmap}. Next, we recall that among $N(N-1)/2$ variables $\mathcal{J}$, there are only $\sim N$ functionally independent ones. Therefore, taking into account $\mathcal{J} = J \cdot \text{sign}(J)$, we can introduce $\sim N$ functionally independent random quasi-spin variables $s_{i}$. Thus, we can rewrite \eqref{XiJform} as follows:
	\begin{equation}
		\Xi^{(int)}_{N}=\left\langle\int e^{ -\tilde{\beta} \sum\limits_{i} J_{i}\,s_{i}}\, \tilde{p}_N (J; \rho) \prod\,dJ_{i}\right\rangle_{s} .
		\label{eq:XiJformchar1}
	\end{equation}
	Suppose that $$ \tilde{p}_N =  \frac{p_N}{\prod\limits_{a}\,p_{1}(\mathcal{J}_a)}$$
	can be interpreted as the density of the conditional probability distribution under a given set of functionally independent interactions $\mathcal{J}_{a}$ considered  as statistically independent. Then the integral over functionally independent $J$ in \eqref{eq:XiJformchar1} can be treated as the analytical continuation of the characteristic function of such a distribution under $\tilde{\beta} \to i\,\tilde{\beta}$, and therefore 
	\begin{equation}
		\Xi^{(int)}_{N}=\left\langle G[s] \right \rangle_{s} .
		\label{eq:XiJformchar2}
	\end{equation}
	The function $G$ has a general structure
	\begin{equation}
		G[s] = \exp\left( 
		\tilde{\beta}\sum_{i}\,h\,s_{i} +  \frac{\tilde{\beta}^2}{2!}\sum_{i,j}\mathfrak{G}_{2} s_{i}\,s_{j}  + \mathcal{O}(\tilde \beta^3)
		\right),
		\label{eq:gspin}
	\end{equation}
	which defines the form of the effective Hamiltonian of the isomorphic Ising-like model, whose quasi-spin couplings generally depend on the density and effective temperature $1/\tilde{\beta}$ related to the temperature of a fluid by \eqref{eq:betabeta}. Note that transforming from particle coordinates to the complete interaction-value distribution is formally exact: provided that all geometrical constraints and correlations are retained in the induced distribution, the transformation preserves the partition function and hence the thermodynamic singularities, phase-diagram topology, and universality class of the original fluid. The resulting generalized quasi-spin Hamiltonian may contain constrained, many-body, and formally nonlocal couplings in the quasi-spin labels, but these encode information already present in the fluid Hamiltonian rather than introducing new physical interactions. For a simple fluid with a single scalar liquid–vapor order parameter, we therefore use the conventional pairwise ferromagnetic Ising Hamiltonian as the minimal model that preserves the essential qualitative features of this ordering sector. The additional effective couplings are not retained because their explicit calculation is prohibitively complicated, and they introduce no additional ordering mechanism relevant to the liquid–vapor transition; their effects are absorbed into the thermal and ordering fields, nonuniversal parameters, analytic background, and corrections to scaling. Thus, the standard pairwise lattice model is a minimal phenomenological realization of the exact interaction representation, rather than a claim that the complete effective Hamiltonian contains only pairwise couplings.	
	\section{Conclusion}\label{sec:end}%
	In this paper, we propose a phenomenological framework for establishing a global isomorphism between continuum fluids and lattice gas (Ising-like) models using a pair of thermodynamic quantities that restore the lattice gas (the Ising model) ``particle-hole`` symmetry between the liquid and gas states of a simple fluid. Such quantities are combinations of density and energy microscopic fields. The description of liquid and gas states, in view of these $\mathbb{Z}_2$-even/odd quantities and their conjugate fields, explicitly reveals a qualitative similarity between these phases. The particle-hole symmetry should now be considered not only in terms of the particle density field but also to include an entropy contribution, i.e., the excluded volume effect. In such a picture, ``particle`` and ``hole`` are indistinguishable. This general idea is represented here in the form of a global isomorphism transformation that maps between the binodals of a continuum fluid and a lattice gas due to the existence of approximate linearity of the density binodal diameter. The crucial test of our approach using the available density and pressure data on a binodal of liquid-gas equilibrium in a 2D fluid, including monolayers of $Ar,Xe, CH_4$, showed fairly good agreement. In the case of monolayers, the honeycomb structure of the graphite substrate seems to support the use of the corresponding lattice model rather than a simple square one, especially when the layer-substrate interaction enhances the ordering according to the substrate lattice structure. However, if we consider the 2D LJ fluid model as representative of the corresponding thermodynamic similarity class based on the critical value of the compressibility factor $Z^{(c)}$ in two and three dimensions, the square lattice Ising-like model with nearest neighbor interactions seems to be an adequate discrete analog. As one can see, the agreement with the data we obtained here is quite satisfactory and much better than in previous attempts to describe liquid-gas equilibrium in 2D fluids based on perturbation methods of liquid theory \cite{crit_2dljpy_jcp1977,eos_2dlj_canjphys1986,crit_yukawa2d_jcp2018}. We outline an exact interaction-variable representation of the fluid partition function and use its sign sector to motivate a generalized quasi-spin description. The conventional pairwise Ising Hamiltonian is adopted as the minimal representative of the liquid--vapor ordering sector: it preserves the essential Ising-like critical structure, whereas the omitted effective couplings contribute to nonuniversal parameters, analytic backgrounds, and corrections to scaling. This construction therefore motivates, but does not constitute a general microscopic proof of, the global projective isomorphism. We believe that this scheme can be tracked analytically using a solvable Kac potential 1D model \cite{crit_kac_physfluid1959} because of the simple linear ordering. The resulting mean-field vdW EoS is a perfect candidate because of the exact linearity of the Zeno-line and the almost straight binodal diameter. Also, we expect that within such a model approach, the relationship between the Zeno-line and the Nishimori line, known in the bond-random Ising model and spin glasses \cite{crit_nishimori_ptp1981}, can be clarified. We will consider these problems in future works.
	\begin{acknowledgments}
	This work was supported by the fortitude of the Armed Forces of Ukraine. V.K. is also grateful to Mr. Konstantin Yun for the financial support of the research. We thank the anonymous reviewer for suggesting that we clarify the relation between the global isomorphism approach and mixed-field scaling.
	\end{acknowledgments}
\end{document}